\documentclass[authoryear,12pt,5p,times]{elsarticle}

\usepackage{bm}
\usepackage{xspace}
\renewcommand{\vec}[1]{\bm{#1}}
\newcommand{\PlP}{\textbf{PlanetPack}\xspace}
\newcommand{\const}{\mathop{\mathrm{const}}}
\newcommand{\disp}{\mathbb{D}}
\newcommand{\FAP}{{\rm FAP}}

\usepackage{amssymb}
\usepackage{amsmath}
\journal{Astronomy and Computing}

\begin{document}
\sloppy

\begin{frontmatter}



\title{\textbf{PlanetPack}: a radial-velocity time-series analysis tool facilitating
exoplanets detection, characterization, and dynamical simulations}


\author{Roman V. Baluev}

\address{Central Astronomical Observatory at Pulkovo of Russian Academy of Sciences,
Pulkovskoje sh. 65, St Petersburg 196140, Russia}
\address{Sobolev Astronomical Institute, St Petersburg State University, Universitetskij
pr. 28, Petrodvorets, St Petersburg 198504, Russia}
 \ead{roman@astro.spbu.ru}

\begin{abstract}
We present PlanetPack, a new software tool that we developed to facilitate and standardize
the advanced analysis of radial velocity (RV) data for the goal of exoplanets detection,
characterization, and basic dynamical $N$-body simulations. PlanetPack is a command-line
interpreter, that can run either in an interactive mode or in a batch mode of automatic
script interpretation.

Its major abilities include: (i) Advanced RV curve fitting with the proper
maximum-likelihood treatment of unknown RV jitter; (ii) User-friendly multi-Keplerian as
well as Newtonian $N$-body RV fits; (iii) Use of more efficient maximum-likelihood
periodograms that involve the full multi-planet fitting (sometimes called as ``residual''
or ``recursive'' periodograms); (iv) Easily calculatable parametric 2D likelihood function
level contours, reflecting the asymptotic confidence regions; (v) Fitting under some
useful functional constraints is user-friendly; (vi) Basic tasks of short- and long-term
planetary dynamical simulation using a fast Everhart-type integrator based on
Gauss--Legendre spacings; (vii) Fitting the data with red noise (auto-correlated errors);
(viii) Various analytical and numerical methods for the tasks of determining the
statistical significance.

It is planned that further functionality may be added to PlanetPack in the future. During
the development of this software, a lot of effort was made to improve the calculational
speed, especially for CPU-demanding tasks. PlanetPack was written in pure C++ (standard of
1998/2003), and is expected to be compilable and usable on a wide range of platforms.
\end{abstract}


\end{frontmatter}

% \linenumbers

\section{Introduction}
\label{sec_intro}
The new era in planetary science has started in 1990s, after the discovery of the first
exoplanet orbiting a main-sequence star \citep{MayorQueloz95}. This discovery was followed
by similar ones in a continuously accelerating regime, and by now the number of known
exoplanets candidates is approaching the notable milestone of thousand (see \emph{The
Extrasolar Planets Encyclopaedia} at \texttt{exoplanet.eu}).

The main method of the exoplanets detection is still the precision radial-velocity (RV)
monitoring. Although with the launch of the specialized spacecrafts like CoRoT
(\texttt{smsc.cnes.fr/COROT}) and Kepler (\texttt{kepler.nasa.gov}) the role of the
photometric searches of exoplanetary transits was considerably emphasized, the RV method
still remains superior in many positions. Even if the discovered transiting exoplanets
will eventually outnumber the ones detected by RV monitoring, the photometric method
introduces a severe bias in favour of the short-period planets. On contrary, the long-term
RV monitoring allows for a detection of exoplanetary systems with architectures resembling
the Solar System, e.g. containing Jupiter analogs~--- giant planets with orbital periods
about a decade or more. Giants with short orbital periods are easier to detect, but they
would hardly allow existence of a terrestial planet on a dynamically stable Earth-like
orbit (although it is still possible to have an Earth-like sattelite of such a gas giant
in the habitable zone). In addition, the RV data are typically necessary to confirm a
photometric exoplanetary detection. From only the transit photometry data we can derive
only the transiter's radius, which does not reliably imply its mass value, and thus
radial-velocity observations are needed to confirm its planetary nature.

Another promising exoplanets detection method is astrometry. It looks relatively latent at
present, but it may become much more productive and efficient in the near future, after
the launch of GAIA. However, we are a bit sceptical about its ability to reliably detect
and characterize long-period exoplanets, because of the relatively short $5$-year expected
duration of the mission. On contrary to space missions, ground-based programmes are able
to accumulate much longer time series. The RV exoplanet searches have already reached the
${\sim} 20$ years baseline.

Therefore, the RV technique is the main tool of exoplanetary seraches at present, and it
will continue to play at least an important, if not central, role in the future. It is
already quite obvious that efficient RV exoplanetary detections request sophisticated
methods of data analysis, which need a specialized software: a good such software complex
is the Systemic Console \citep{Meschiari09}. Our paper represents a scientific description
of another such software tool that we developed for similar same goals. The need for
another software tool was justified by the following argumentation:
\begin{enumerate}
\item Systemic Console relies on rather simple statistical methods and models that appear
inadequate when working with high-precision exoplanetary RV data. For example, it relies
on the plain $\chi^2$ fitting and on the textbook $F$-test, which are unreliable for the
RV noise appearing in our task \citep{Baluev08b}. We needed to implement some more
intricate statistical treatment, especially in what concerns periodograms.

\item Systemic Console was written in JAVA to reach cross-platform compatibility, but this
leaded to a dramatic decrease in the computational performance, which appears pretty
obvious when working with Systemic. In scientific tasks the speed of calculations is
usually a more important matter than the wide compatibility. \PlP is written in standard
C++, and thus it is quick. It should be easily compilable by different compilers and for
various platforms, although it was extensively tested only with the GCC and Linux-based
environments.

\item It appears that Systemic Console is targeted to amateur astronomers: e.g.\ it is more
focused on the graphical interface rather than on a dense scientific content. We needed to
focus mainly on the scientific contribution and more scripting capabilities, shifting the
basis of the software to a command-line interface, since it allows for a more controllable
and powerful work environment.

\item According to the information in the official Systemic web page at oklo.org, this
package was last updated in 2009.
\end{enumerate}

A few more recent software tools intended for exoplanetary data fits are available today.
In particular, \citet{WrightHoward09} provide an algorithm of exoplanets RV fitting,
taking into account the fact that there are a few strictly linear parameters of Keplerian
RV variation, that can be efficiently eliminated during the fitting of the remaining
non-linear parameters. This algorithm assumes that the gravitational perturbations in the
exoplanetary system are negligible. \citet{Pal10} provided an RV fitting algorithm for the
self-perturbed exoplanetary systems, based on the Lie integration scheme. And finally,
\citet{Eastman13} offer an algorithm of simultaneous ``photometry+RV'' fitting, also
equipped by some Bayesian Markov chain Monte Carlo simulation tools.

We have not done a ``field-test'' performance comparison of these packages with \PlP.
Nevertheless, we may note that \PlP offers some highly importaint algorithms that are
unavailable in other packages (in particular, the red-noise RV fitting and the advanced
periodograms construction) and, on contrary, the mentioned packages offer some important
tools that are absent in \PlP (in particular, the joint analysis of photometry and radial
velocities, Bayesian statistics). The practical value of our new package, as we see it, is
in its wide task coverage: it unites under the same umbrella a large number of very
different particular tools in a single place. When developing \PlP the main effort was
done in the direction of the data-analysis methods, rather than just in programming or
optimizing computational performance. Almost all data-analysis methods that \PlP
incorporates belong to the self-consistent theory work that we carried out over a few
years.

This was not just a pure theoretical investigation: we applied our tools to real
exoplanetary systems, so that these data-analysis methods were evolving and improving in
this process. Moreover, this allowed to obtain new results concerning the relevant
exoplanetary systems, and most of these concrete results were eventually confirmed by
independent authors, often based on the enlarged and/or improved datasets. Such examples
include the rejection of the planet HD74156~\emph{d} (disclaimed by \citet{Baluev08b},
further retracted by \citet{Wittenmyer09} and \citet{Meschiari11}); the revealing of the
2/1 resonance in the HD37124 planetary system (first revealed in \citet{Baluev08c}, later
confirmed by \citet{Wright11}), and the detection of the hints of the planet
GJ876~\emph{e} (as we discuss in \citet{Baluev08-IAUS249,Baluev11}, a good and stable
orbit for this planet can be found in the old RV data, long before its announcement by
\citet{Rivera10}). We draw the reader's attention to these examples not for bragging,
but in order to highlight the potential of the theory and ideas that we collect now under
the name ``\PlP''. The demonstrated examples prove that this software tool may
significantly increase the outcome of the ongoing exoplanetary RV data-analysis work, as
well as to prevent us from too hasteful conclusions.

\PlP source, along with its technical manual, is available for download as a project at
\texttt{sourceforge.net/projects/planetpack}. In the further sections of the present
paper, we consider the main \PlP abilities and the related theory. This paper does not say
anything about the use of \PlP commands, its data organization, and other technical
documentation necessary to use it in practice. The mentioned technical documentation is
given in a standalone file downloadable together with
\PlP sources.

\section{Data and basic models}
\label{sec_datamod}
Let us first describe the general structure of the observational data set that we deal
with. Assume that we have $J$ RV time-series, referring to the same star but to different
observatories or spectrographs. A $j$-th such time series contains $N_j$ elementary data
packets, consisting of the time of an observation, $t_{ji}$, of the RV measurement itself,
$v_{ji}$, and of its expected uncertainty $\sigma_{\mathrm{meas},ji}$. The total number of
these observations is $N=\sum_{j=1}^J N_j$.

In addition to this raw input data, \PlP uses a time reference epoch, $T_0$, as an
unfittable parameter. Before any fitting, the values of $t_i$ are always shifted by this
quantity and divided by the total time-series span, $T$ (calculated internally).
Therefore, the values that are actually used are $(t_i-T_0)/T$. This process should
normally remain invisible to the user, but to minimize numerical errors, it is recommended
to choose $T_0$ close to the middle of the time series. This $T_0$ is also used as a
reference epoch for the orbital parameters, when such a reference epoch is necessary (see
below). The desired value of $T_0$ can be assigned explicitly by the user or it may be
chosen automatically (a round number close to the weighted mean of $t_i$). Below we assume
that $T_0=0$ for the simplicity of the formulae. The transition to the case of $T_0$ is
obvious.

Now let us specify the general functional model of the RV curve. It is basically the same
as we used in \citet{Baluev08c}. For each of the $J$ time series we have a separate model
that can be represented as the following sum:
\begin{equation}
\mu_j(t,\vec\theta) = \mu_{\mathrm{obs},j}(t,\vec\theta_{\mathrm{obs},j}) +
\mu_\star(t,\vec\theta_\star).
\label{RVmod}
\end{equation}
This is a sum of two terms. The first term, $\mu_{\mathrm{obs},j}$, depends on the time
series through the index $j$, and it represents an observatory-specific part of the
measured radial velocity:
\begin{equation}
\mu_{\mathrm{obs},j} = c_{0,j} +
  \sum_{n=1}^{s_j} A_{jn} \cos\left(\frac{2\pi}{P_{jn}}(t-\tau_{jn})\right).
\label{RVmod_obs}
\end{equation}
In this definition, the term $c_{0,j}$ is a constant term denoting an RV offset of the
$j$-th time series, and the remaining (periodic) terms model possible observatory-specific
periodic components, e.g.\ systematic errors. The compound vector
$\vec\theta_{\mathrm{obs},j}$ contains the variables $c_{0,j}$, $A_{jn}$ (the
semi-amplitude of a systematic term), and $\tau_{jn}$ (the epoch of the maximum systematic
variation, treated relatively to $T_0$). The periods $P_{jn}$ are treated as fixed
parameters.

We may recall that e.g.\ annual systematic errors can be rather frequently met in the
published exoplanetary RV data, especially in the old datasets, where they may exceed
${\sim} 10$~m/s \citep{Baluev08c,Baluev08b}. Although this our conclusion was first
considered with some scepticism by other researchers, at present such errors have been
revealed by independent teams \citep{Wittenmyer09,Meschiari11} and sometimes they can be
clearly and undoubtfully seen when comparing published old and revised RV data for the
same star \citep{Baluev11}. We believe that the existence of such errors in some of the
publicly released RV data of exoplnetary systems is already proven well. Although we must
admit that in recent years the major observing teems seem to do a good job on removing
this issue, the old data, which are certainly useful, may still suffer from such errors.
Therefore \PlP still allows to deal with this issue by means of an expanded
model~(\ref{RVmod_obs}).

The second term in~(\ref{RVmod}) is common for all time series; it referes to the star and
its planetary system and it has the general form of
\begin{equation}
\mu_\star(t,\vec\theta_\star) = \sum_{n=1}^r c_n t^n + \mu_\mathrm{pl}(t,\vec\theta_\mathrm{pl}),
\end{equation}
where $c_n$ are coefficients of a polynomial trend modelling some long-term underlying RV
variation (usually it reflects the compound RV contribution from some long-period seen or
unsees bodies in the system), and $\mu_\mathrm{pl}$ describes the RV variation due to the
assumed orbiting exoplanets (each with an individual and independent RV contribution). The
vector $\vec\theta_\star$ contains the coefficients $c_n$ and the elements of
$\vec\theta_\mathrm{pl}$. Notice that $c_n$ are understood in view of the reference epoch
$T_0$.

The first published version of \PlP may set only a \emph{common} polynomial trend for the
whole time series. Sometimes it might be useful to allow for separate datasets to have
different trends, reflecting e.g. some long-term instrumental drifts. This ability was not
implemented in \PlP till now, because we have not yet faced a practical task where this
would be necessary, but this may be done in the future. At present, the models with
different trends may already be constructed with a help of a ruse: to obtain, e.g. an
almost quadratic trend in the model of only some specific dataset, we need to specify in
the relevant sum~(\ref{RVmod_obs}) a harmonic term with a very large period value (larger
than the observations time span). A linear trend can be mimiced by means of setting a
constraint (Sect.~\ref{sec_con}) to fix one of the two parameters of this long-period
harmonic term.

In the simplest and most frequent case, when the interplanetary gravitational
perturbations in the system are negligible, we may assume the multi-Keplerian model
\begin{equation}
\mu_\mathrm{pl} = \sum_{k=1}^{\mathcal N} K_k (\cos(\omega_k+\upsilon_k)+e_k\cos\omega_k).
\end{equation}
Here $\mathcal N$ is the number of orbiting exoplanets, $K_k$ is the RV semi-amplitude
induced by $k^\mathrm{th}$ exoplanet, $e_k$ is the relevant orbital eccentricity,
$\omega_k$ is the pericenter argument, and $\upsilon_k$ is the true anomaly. The true
anomaly can be represented as a function of the time $t$, of the mean-motion $n_k$, of
$e_k$, and of an additional phase parameter $\lambda_k$. We choose this phase parameter to
be the mean longitude at $T_0$. Therefore, the vector $\vec\theta_\mathrm{pl}$ contains
the variables $(n,K,\lambda,e,\omega)_k$ for each of the $\mathcal N$ planets. Notice that
for an exoplanet on a circular orbit we have the relevant RV variation looking like $K_k
\cos(n_k t + \lambda_k)$.

For some time, we investigated the possibility to fit the parameters $e\cos\omega$ and
$e\sin\omega$ instead of $e$ and $\omega$, since the last pair implies an undesired
singularity at $e=0$. However, we did not note any increase in the fitting performance
after the transition to $(e\cos\omega,e\sin\omega)$. Moreover, it appeared that in the
practical tests the resulting convergence rate actually dropped after that transition, and
even for small-eccentricity orbits. We therefore abandoned this idea and returned to the
direct fitting of $(e,\omega)$. However, when $e$ is small, the user should be careful
with the interpretation of its uncertainty reported by \PlP: in this case, the uncertainty
of $e$ becomes meaningless without an accompanying uncertainty of $\omega$ and without the
correlation between $e$ and $\omega$. Actually, in this case the best course of action
would be to look at the 2D confidence contours (Sect.~\ref{sec_confreg}) plotted in the
plane $(e\cos\omega,e\sin\omega)$, assuming that $e$ and $\omega$ are polar coordinates.
Such a plot would be much more informative in this case, than e.g. just an upper limit on
$e$.

The minimum mass of an exoplanet, $m\sin i$, and the semi-major axis of its orbit, $a$,
can be expressed via the primary fit parameters using the well-known relations
\begin{eqnarray}
m \sin i \simeq \tilde K \left(\frac{M_\star^2}{G n}\right)^{1/3} = \mathcal M \tilde K M_\star^{2/3} n^{-1/3}, \nonumber\\
a \simeq \left(\frac{G M_\star}{n^2}\right) = \mathcal A M_\star^{1/3} n^{-2/3},
\label{ma}
\end{eqnarray}
where $\tilde K = K \sqrt{1-e^2}$, $G$ is the gravitational constant, and $M_\star$ is the
mass of the star (which should be derived from some external considerations), and
$\mathcal M$ and $\mathcal A$ are conversion constants ($\mathcal M \approx 9.077\cdot
10^{-3}$ and $\mathcal A \approx 6.664\cdot 10^{-2}$ when the unit of $m$ is
$M_\mathrm{Jup}$, of $M_\star$ is $M_\odot$, of $n$ is day$^{-1}$, and of $\tilde K$ is
m/s). \PlP uses $\tilde K$ as primary parameter instead of $K$, since then its conversion
to $m\sin i$ does not involve the eccentricity $e$ (which also eliminates the need to take
into account the corelation with $e$ when evaluating the uncertainty of $m\sin i$).

The approximate formulae~(\ref{ma}) are valid when $m\ll M_\star$, which is true for the
most practical cases. More accurate formulae, which take into account barycenter effects,
exist \citep{Ferraz-Mello-lec1,Pal10,Beauge12} and are rather popular in practice.
However, for multi-Keplerian firts we do not accept this approach due to the following
reasons:
\begin{enumerate}
\item These formulae are implicit and therefore more difficult for practical use.
\item They involve significant dependence on the orbital inclination (the famous $\sin
i$), which is typically unknown. Eventually we have to assume e.g.\ $i=90^\circ$, and if
this assumtion is wrong, the ``corrected'' mass value will anyway contain a remaining
error comparable to the original one.
\item They are not actually more accurate than~(\ref{ma}), unless we deal with a
single-planet system. When the system contains two or more planets we should also take
into account mutual gravitational perturbations, including e.g.\ the offset in the
apparent period value \citep{Ferraz-Mello-lec1}, which would affect the resulting mass
value too. These biases of the order $m/M_\star$ are typically neglected, but then there
is no reason to take into account any other terms with a similar magnitude, including
those due to the barycenter displacement.
\item For the unperturbed exoplanetary case the formulae~(\ref{ma}) are more than
satisfactory, because the errors due to statistical uncertainties are dominating anyway.
\end{enumerate}
We may note that in the case of the Newtonian $N$-body fitting (Sect.~\ref{sec_nbody}),
\PlP will honestly evaluate the correct planet masses, taking into account all
gravitational effects and the best-fit value of $\sin i$. This is achieved using an
artificial ``osculating RV semi-amplitude'' parameter, see the details in
\citep{Baluev11}. Also, we would like to emphasize that the primary fit parameters are $K$
and $P$, not $m$, and the formula used to obtain $m$ do not affect the fitting process in
any way. It only affects the value of $m$ derived \emph{after} the fit.

\PlP deals with the parametrized RV noise. The basic noise model assumes that the errors
of all $v_{ji}$ are independent and Gaussian with the variances expressed as
\begin{equation}
\sigma_{ji}^2 = \sigma_{\mathrm{meas},ji}^2 + \sigma_{\star,j}^2,
\end{equation}
where the quantities $p_j=\sigma_{\star,j}^2$ represent additional unknown parameters (RV
``jitter'') to be estimated from the data. These parameters can be combined in a single
vector $\vec p$. Notice that we understand $\sigma_{\star,j}^2$ as a solid symbol here,
because in practice we may sometimes deal with the cases $p_j<0$, meaning that the values
of $\sigma_{\mathrm{meas},ji}$ supplied by the observers possess rather poor quality, and
the real errors of $v_{ji}$ are systematically smaller \citep{Baluev08b}. As we have
already discussed in that paper, in practice the apparent RV jitter often have little
resemblance with the actual RV instability caused by astrophysical effects on the star
itself. The instrumental errors and various spectrum reduction imperfections may introduce
a comparable and even dominating contributions. We should treat $p_j$ just as free
parameters introduced to reach some degree of model consistency, avoiding to assign any
concrete physical sense to them.

\section{Maximum-likelihood RV curve fitting}
Assuming the uncorrelated Gaussian distribution of RV errors, we can write down the
likelihood function of the task as
\begin{eqnarray}
\ln \mathcal L(\vec\theta,\vec p) = -\frac{1}{2} \sum_{j=1}^J \sum_{i=1}^{N_j} \left[ \ln\sigma_{ji}^2(\vec p) + \phantom{\frac{\left(v_{ji}\right)^2}{\gamma \sigma_{ji}^2}} \right. & & \nonumber\\
\left. +\frac{\left(v_{ji}-\mu_j(t_{ji},\vec\theta)\right)^2}{\sigma_{ji}^2(\vec p)} \right] - \frac{N}{2} \ln{2\pi}. & &
\label{loglik}
\end{eqnarray}
The position of the maximum of~(\ref{loglik}) would yield the classic maximum-likelihood
estimation of the parameters $\vec\theta$ and $\vec p$. However, \PlP uses a modification
of the maximum-likelihood method, which is based on the following modified likelihood
function:
\begin{eqnarray}
\ln \tilde{\mathcal L}(\vec\theta,\vec p) = -\frac{1}{2} \sum_{j=1}^J \sum_{i=1}^{N_j} \left[ \ln\sigma_{ji}^2(\vec p) + \phantom{\frac{\left(v_{ji}\right)^2}{\gamma \sigma_{ji}^2}} \right. & & \nonumber\\
\left. +\frac{\left(v_{ji}-\mu_j(t_{ji},\vec\theta)\right)^2}{\gamma \sigma_{ji}^2(\vec p)} \right] - \frac{N}{2} \ln{2\pi}. & &
\label{loglikmod}
\end{eqnarray}
The best-fitting estimations of $\vec\theta$ and $\vec p$ are obtained as the position of
the maximum of $\ln \tilde{\mathcal L}$. The thing that makes the
definition~(\ref{loglikmod}) to differ from the classic one in~(\ref{loglik}) is the
correction divisor $\gamma$. It is equal to $\gamma=1-d/N$, where $d$ is the number of the
degrees of freedom of the RV model, here equal to $\dim\vec\theta$. The purpose of the
corrector $\gamma$ is to reduce the systematic bias in the estimation of $\vec p$ that
would otherwise appear due to the fact that the best-fit residuals are systematically
smaller than the actual measurement errors. See the details in \citep{Baluev08b}.

The larger is the resulting maximum value of $\tilde{\mathcal L}$, the better is the fit
quality. The value of $\tilde{\mathcal L}$ is not very intuitive, however. As a numerical
measure of the fit quality we offer a more useful quantity
\begin{equation}
\tilde l = 0.2420 \tilde{\mathcal L}^{-1/N},
\end{equation}
because it is resembling the traditional r.m.s. measure. First, the smaller is $\tilde l$,
the better is the fit. Second, $\tilde l$ is measured in the same units as the
observations $v_{ji}$ (i.e., in m/s). And third, the normalization of $\tilde l$ is chosen
so that $\tilde l$ approximately reflects an average of the RV residuals.

The detailed theory and justification of this method is given in \citep{Baluev08b}. \PlP
performs the non-linear maximization of~(\ref{loglikmod}) using a variant of the
Levenberg--Marquardt (LM) algorithm. Our implementation of this algorithm is different
from e.g.\ the one used in the wide-spread MINPACK library, because the latter was
designed to deal with only a sum-of-squares objective function, emerging in the
least-squares regression task. This special structure of the objective would allow to use
certain simplifying relations between its gradient and the Hessian matrix, but our
objective~(\ref{loglikmod}) does not belong to this class. Although we describe in
\citep{Baluev08b} a way to ``fool'' the MINPACK or MINPACK-like algorithms, forcing them
to solve the task we actually need, we eventually decided to use our own variant of the LM
algorithm, more general than the one used in MINPACK. Our implementation represents some
hybrid method between the MINPACK variant and the classic general one described in
\citep{Bard}. It does not rely on the assumption that the objective is a sum of squares.

\section{Advanced periodograms}
\label{sec_prdg}
\PlP is equipped with improved versions of the periodograms, which have many advantages in
comparison with the classic \citet{Lomb76}--\citet{Scargle82} periodogram. Their main
improvements are listed below.
\begin{enumerate}
\item These periodograms are the likelihood-ratio periodograms. Their values basically
represent the likelihood-ratio statistic associated with the modified likelihood
function~(\ref{loglikmod}). The motivation and details of this approach are given in
\citep{Baluev08a,Baluev13b}. In particular, such periodograms involve a built-in
estimation of the RV jitter and other RV noise parameters, which allows for a
self-consistent data fitting already at the period-search stage.

\item At the very beginning of the analysis, these periodograms can used to just detect a
periodic signal in a raw input time series. But they may also be used in further steps,
when one or a few planets have been already extracted from the data, and we need to check
whether the residuals hide an additional planet. However, these periodograms are not just
the plain periodograms of the relevant pre-calculated and then frozen residuals, as it is
typically done in this task. \PlP evaluates \emph{each} value of such a periodogram by
means of a full multi-planet fit, which is performed almost \emph{anew}, re-adjusting
e.g.\ the parameters of already extracted planets. The advantage of such periodograms is
clearly demonstrated by \citet{Anglada-Escude12}, who call them ``recursive
periodograms''. We prefer to call them as the ``residual periodograms'', on contrary to
the ``periodograms of residuals''. This can also be treated as a broad extension of the
generalized ``floating-mean'' periodogram \citep{FerrazMello81,Cumming99,ZechKur09}.

\item The \PlP's periodograms can utilize the simple sinusoidal model of the signal, as
well as a more complicated periodic model representing a segment of the Fourier series
(trigonometric polynomial of a given degree). The periodogram with the sinusoidal model
represents an extension of the classic Lomb--Scargle periodogram, while the periodogram
with the Fourier model is a similar extension of the so-called multiharmonic periodogram
\citep{SchwCzerny96,Baluev09a}. The Fourier model may be more suitable for non-sinusoidal
RV signals, which may appear due to planets on highly-eccentric orbits.
\end{enumerate}

Therefore, the individual values of the \PlP's periodograms actually represent the
modified likelihood-ratio statistic $\tilde Z$ of Section~\ref{sec_stat} below. The base
RV model describes our knowledge of the planetary system at the current step of the
analysis, while the alternative one also involves a trial periodic signal modelled by a
sinusoid or trigonometric polynomial (having a given basic period). The issues related to
the stiatistical significance levels of these periodograms will be discussed in detail in
Section~\ref{sec_stat}.

\section{Constrained fitting}
\label{sec_con}
\PlP allows to perform the maximum-likelihood fitting under some simple equality
constraints. Let us denote full vector of the RV curve parameters, consisting of the RV
curve parameters $\vec\theta$ and of the noise parameters $\vec p$, as $\vec\xi$. Let us
assume that we need to maximize~(\ref{loglikmod}) under a condition $\vec\eta(\vec\xi) =
\vec\eta_0$, where $\vec\eta$ is a specified vector function of a vector argument, and
$\vec\eta_0$ is a vector constant. In this case we need to find
\begin{eqnarray}
\tilde{\mathcal L}^*(\vec\eta_0) &=& \left. \max_{\vec\xi} \tilde{\mathcal L}(\vec\xi)\right|_{\vec\eta(\vec\xi)=\vec\eta_0}, \nonumber\\
\vec\xi^* (\vec \eta_0) &=& \left. \arg \max_{\vec\xi} \tilde{\mathcal L}(\vec\xi)\right|_{\vec\eta(\vec\xi)=\vec\eta_0}.
\label{likmaxeta}
\end{eqnarray}

At present there is only rather limited, though useful, set of functions that can be
chosen as constraints. Namely, it is allowed to constrain any single fit parameter (either
primary or derived one, including the amplitudes $K$ and $\tilde K$, and the minimum mass
$m\sin i$), a mutual inclination between planetary orbits (in the case when it appears
constrainable from the RV data thanks to the gravitational perturbations), and the mutual
inclination with an accompanying nodes line orientation angle (see \citealt{Baluev11} and
\PlP manual for further details).

The procedure of the constrained fitting when one or more primary fit parameters are held
fixed is trivial: we just need to ignore the relevant parameters in the LM algorithm. When
a combination of two or more primary parameters is constrained, we use the method of
elimination to perform this constrained fitting. That is, during the fitting we directly
express some of the parameters involved in $\vec\eta$ via the remaining ones by means of
explicit formulae, and also adjust the gradient and the Hessian approximation for
$\tilde{\mathcal L}$ to take this elimination into account.

Notice that the constraint in~(\ref{likmaxeta}) implies a decrease in the number of
degrees of freedom of the RV model, which affects the value of the corrector $\gamma$
in~(\ref{loglikmod}). In the constrained case we have
$\gamma=1-(\dim\vec\theta-\dim\vec\eta)/N$, provided that all constraints in $\vec\eta$
refer to the RV curve model (the RV noise parameters, as well as any their constraints, do
not affect $\gamma$).

\section{Parametric confidence regions}
\label{sec_confreg}
\PlP makes it easy to construct the level contours of the function~(\ref{loglikmod}),
which can serve as asymptotic parametric confidence regions. The method is generally
similar to the one described in \citep{Baluev08c}. Let our full vector of the RV curve
parameters be $\vec\xi$, and we need to to construct the confidence region for the
variables $\vec\eta=\vec\eta(\vec\xi)$. This new vector $\vec\eta$ has necessarily smaller
dimension than $\vec\xi$ (in practice usually there are only one or two parameters in
$\vec\eta$) and it may represent just a subset of $\vec\xi$ or some simple function of
$\vec\xi$ (among those described in Section~\ref{sec_con}). Then, for a given trial
$\vec\eta_0$ from a multi-dimensional grid, we perform the following constrained
fitting~(\ref{likmaxeta}).

The partly-maximized function $\tilde{\mathcal L}^*$ in~(\ref{likmaxeta}) can be plotted
on a multi-dimensional grid of $\vec\eta$, and its level contours will represent the
necessary confidence regions. We need to notice that \PlP does not contain any graphical
plotting facilities; it only generates a table of the quantities $\vec\eta$,
$\tilde{\mathcal L}^*(\vec\eta)$, $\vec\xi^*(\vec\eta)$, which is supposed to be used
later by an external graphical plotter (like e.g.\ GNUPLOT).

We still need to calibrate these level contours with the actual significance probability.
For this goal, we also need to define the following quantities, produced by the usual
unconstrained fitting:
\begin{eqnarray}
\tilde{\mathcal L}^{**} &=& \max_{\vec\xi} \tilde{\mathcal L}(\vec\xi) , \nonumber\\
\vec\xi^{**} &=& \arg \max_{\vec\xi} \tilde{\mathcal L}(\vec\xi).
\label{likmax}
\end{eqnarray}
Then, following \citep{Baluev08b}, we can pose a hypothesis testing task, with the
encompassing (alternative) hypothesis $\mathcal K$: ``$\vec\xi$ is arbitrary'' (implying
the best-fitting estimation $\vec\xi = \vec\xi^{**}$ and $\tilde{\mathcal L}_{\mathcal K} =
\tilde{\mathcal L}^{**}$), and the restricted (base) hypothesis $\mathcal H$: ``$\vec\xi$
satisfies the constraint $\vec\eta(\vec\xi) = \const$'' (implying $\vec\xi =
\vec\xi^*(\vec\eta)$ and $\tilde{\mathcal L}_{\mathcal H} = \tilde{\mathcal
L}^*(\vec\eta)$). The numbers of the degrees of freedom in the relevant models are now
$d_{\mathcal H} = \dim\vec\xi-\dim\vec\eta$ and $d_{\mathcal K} = \dim\vec\xi$. Note that
due to the divisor $\gamma$ in~(\ref{loglikmod}), which depends on the number of free
parameters (hence, on the number of constraints too), the function $\tilde{\mathcal L}$ is
a bit different in~(\ref{likmaxeta}) and in~(\ref{likmax}). This means, e.g., that
$\tilde{\mathcal L}^{**} \neq \max_{\vec\eta} \tilde{\mathcal L}^*(\vec\eta)$ in our
case.

The confidence level for a given likelihood contour $\tilde{\mathcal L}^*(\vec\eta) =
\const$ can be mapped with the relevant likelihood-ratio statistic $\tilde Z$ of the below
Section~\ref{sec_stat}, with $\tilde{\mathcal L}_{\mathcal H} = \tilde{\mathcal L}^*$ and
$\tilde{\mathcal L}_{\mathcal K} = \tilde{\mathcal L}^{**}$. From the well-known classical
results it follows that when $N\to\infty$, the quantity $2\tilde Z$ asymptotically follows
the $\chi^2$ distribution with $d=d_{\mathcal K}-d_{\mathcal H}=\dim\vec\eta$ degrees of
freedom.\footnote{Possible constraints set on the parameters of the RV \emph{noise}
(rather than \emph{curve}) model, do not affect the corrector $\gamma$, but still affect
the asymptotic $\chi^2$ distribution of $\tilde Z$, increasing its number of degrees of
freedom. \PlP is aware of this subtle issue and deals with it properly.} Therefore, the
overall sequence to obtain the asymptotic confidence regions for the parameters $\vec\eta$
looks like the following:
\begin{enumerate}
\item Obtain the necessary table of $\tilde{\mathcal L}^*(\vec\eta)$.
\item For a selected contour value $\tilde{\mathcal L}^*$ and pre-calculated
$\tilde{\mathcal L}^{**}$, construct the statistic $\tilde Z$.
\item Evaluate the corresponding asymptotic $\chi^2$ confidence probability as $P_d(\tilde
Z)$, where $P_d(z) = \left. \Gamma_z\left(\frac{d}{2}\right)\right/
\Gamma\left(\frac{d}{2}\right)$, with $\Gamma_z$ being the incomplete gamma function.
\end{enumerate}
The step two is actually done by \PlP automatically: the relevant value of $\tilde Z$ is
written in the output table along with the other data. The probability $P_d(\tilde Z)$ is
not evaluated by \PlP, because it would require an access to non-standard math libraries,
which we try to avoid. However, the necessary gamma function is available in GNUPLOT,
which we recommend to use when plotting the relevant probability contours in a graph.

\section{Red-noise analysis}
\label{sec_rednoise}
\PlP can deal with the RV data contaminated by the correlated (``red'') noise. This red
noise appears rather frequently in practice and imposes a lot of misleading effects
without proper treatment \citep{Baluev11,Baluev13a}. \PlP uses the maximum-likelihood
algorithm of the red-noise reduction, which is described in \citep{Baluev13a} in full
details. The RV noise model is now more complicated than the basic white-noise one
described in Section~\ref{sec_datamod}. It is modelled by a Gaussian random process with
an exponentially decreasing correlation function. In the white-noise case we had only a
single noise parameter, $\sigma_\star^2$ (or a few such parameters for different time
series). The free parameters of the correlated RV noise model are: the variance of the
``white'' noise component $\sigma_\mathrm{white}^2$, the variance of the ``red'' noise
component $\sigma_\mathrm{red}^2$, and the noise correlation timescale
$\tau_\mathrm{red}$, such that the covariance coefficient between two RV measurements
separated by the time gap $\Delta t$ is equal to $\sigma_\mathrm{red}^2 e^{-\Delta
t/\tau_\mathrm{red}}$. This parameter $\tau_\mathrm{red}$ should not be mixed with
$\tau_{jn}$ from~(\ref{RVmod_obs}).

\PlP allows for two types of the red-noise models: the model with a ``shared'' red noise
and with a ``separated'' red noise. In the shared model, \PlP deals with only a single
pair of the red-noise parameters $(\sigma_\mathrm{red}^2,\tau_\mathrm{red})$, and the red
component of the noise is the same for all RV data points of the joint time series. This
model means that the red noise is generated by the star itself, and not by individual
instruments. The white parts of the noise are still assumed different for individual data
sets of the compound time series. In the second, separated, model, the red noise is
treated separately for to each individual datasets, and the number of the red-noise
parameters is increased accordingly. The correlations between RV data points belonging to
different datasets (i.e., different spectrographs) are set to zero in such a case. It is
also possible to specify red-noise component to only some of the datasets, leaving the
others purely white. It is not allowed to specify one or more separated red-noise
component when a shared red noise is defined, because such a model would be very close to
the degeneracy.

Although this method of red-noise reduction is rather new, it have proven its high
practical efficiency in the case of the exoplanetary systems of GJ876 \citep{Baluev11} and
GJ581 \citep{Baluev13a}. We believe that it should appear useful in other cases too, so we
offer its implementation in \PlP.

\section{Newtonian $N$-body fitting and dynamics}
\label{sec_nbody}
Some exoplanetary systems show detectable hints of non-Keplerian dynamics due to
interplantery perturbations. In this case a more complicated RV curve model should be
used, which should be based on the numerical integration of the relevant $N$-body task.
The algorithm of $N$-body fitting used by \PlP is the one described in \citep{Baluev11} in
all details. This algorithm involves an integration of the $N$-body equations for the
planetary coordinates and velocities together with the associated differential equations
for their partial derivatives with respect to the osculating orbital elements (i.e., the
variational or sensitivity equations). This method allows to calculate the necessary
objective function, its gradient, and its Hessian matrix with much better speed/error
ratio then e.g.\ evaluating the gradient via finite differences. The osculating orbital
parameters are referenced in the Jacobi coordinate system. Please see \citep{Baluev11} for
the explanation of the method, coordinate system, and other details.

The choice of the Jacobi system is motivated by the fact that it allows much more smooth
switching between perturbed and unperturbed RV models. The main difficulty in such a
transition comes from the planetary orbital period estimaions: the \emph{apparent} period
(the one seen in an RV periodogram) is different from the \emph{osculating} period. The
first-order formula for this displacement is given in \citep{Ferraz-Mello-lec1}. This
offset appears due to the secular perturbations in the planetary mean longitude, and it
has the following bad consequence. Assume we performed a non-perturbed fit and found a
best-fit (apparent, or periodogram) period for some planet. After that, we may wish to see
what will change if we add the interplanetary perturbations. When making a perturbed fit,
we have no other option than to treat the apparent period value as the osculating one, but
these values are different by their definition! In other words, feeding the $N$-body model
with the observed period value generates a biased actual (averaged) model period; it is
displaced from the observed period that we have just substituted as the osculating one. In
the worst cases, we may discover that our fitting algorithm refuses to converge to
anything reasonable at all, because the relevant frequency displacement exceeds (even
significantly exceeds) the periodogram resolution $\sim 1/T$. This was the case for the
planet GJ876~\emph{d} in \citep{Baluev11}, for example. Ideally, we should first reduce
this displacement between the periods, e.g.\ according to the formulae by
\citet{Ferraz-Mello-lec1}. However, in our previous works we have established (rather
empirically) that the use of the Jacobi coordinates practically eliminates the need of
such a period correction: the osculating Jacobi periods appear much closer to the apparent
periods, than the osculating periods referenced in other coordinate systems. Such an
effect is achieved because we refer the osculating orbital periods to an increased star
mass value, incorporating the mass of the planet, whose osculating period we want to
define, and also of the planets below it (among those included in the integration). Again,
see \citep{Baluev11} for the details.

To make such $N$-body fitting to work we obviously need a numerical integrator. \PlP uses
an extension of the old \citet{Everhart74} integrator for this goal. As it was discussed
by \citet{Avdushev10}, the Everhart integrator is, basically, an implicit Runge-Kutta
integrator, equipped by an efficent predictor evaluation. The original Everhart integrator
was based on the Gauss--Radau or Gauss--Lobatto splitting of each integration step.
\citet{Avdushev10} gives the general formulae suitable for an arbitary sequence of the
splitting nodes. In particular, the Gauss--Legendre and Gauss--Lobatto spacings generate an
integrator with a useful symplectic property (when the integration step is constant). The
integrator used in \PlP is an $16$-th order integrator, based on $8$ Gauss--Legendre nodes.
In comparision with the \citet{Avdushev10} implementation, we introduced some changes to
increase the calculation speed:
\begin{enumerate}
\item The formulae given in \citep{Avdushev10} are valid for a system of first-order
equations $\dot{\vec x}=F(\vec x)$. We extended them to the second-order case which we
actually deal with, $\ddot{\vec x}=F(\vec x)$. This allowed to increase the integrator
performance roughly twice, in comparison with the trivial substitution $\vec y = \{\vec
x,\dot{\vec x}\}$ leading to the first-order system $\dot{\vec y} = G(\vec y) =
\{\dot{\vec x}, F(\vec x)\}$. The necessary corrections are fairly obvious when comparing
the formulae of the original Everhart method with the general formulae by Avdushev. We do
not detail these changes here, since this would require to replicate a large part of the
Avdushev's paper.

\item On contrary to \citep{Avdushev10}, in the source code we define the integration
nodes as compile-time constants. All other derived coefficients and constants of the
scheme are pre-calculated at the compilation stage as well (i.e., before the execution of
\PlP itself). This also improves the calculation speed significantly. However, this goal
was reached by means of rather sophisticated template metaprogramming tools of C++, which
requires a fully standard-compliant compiler with a clever code optimization. For example,
GCC or Intel C++ Compiler work well (when proper optimization options are turned on),
while with MS Visual C++ compiler we failed to achieve the same fast code.

\item The step-size control method of the original Everhart integrator is imperfect. If
$s$ stands for the number of integration nodes, the step size is adjusted as if the
integrator had the order of $s$, but for the specific node systems like, e.g., Lobatto,
Radau, or Legendre ones, the actual integrator order is equal to $2s-2$, $2s-1$, or $2s$,
respectively. In case of our $16$-th order integrator, the step would be chosen in a very
pessimistic manner, as if the integration order was only $8$. Then the resulting
integration errors would be much smaller than what we request. We have established
empirically that the actual integration error appears roughly equal to the square of the
requested one. Therefore, we correct the step-size control procedure by passing the
\emph{square root} of the desired relative precision, instead of the desired relative
precision itself. In practice this simple method works nicely: the step is scaled
according to the actual integrator order ($16$), and the actual precision of the
integrator is in much better agreement with the requested one ($1-2$ orders in magnitude).
\end{enumerate}

All these changes leaded to a significant cumulative increase in the speed of the
calculations, in comparison with the FORTRAN code provided by \citet{Avdushev10}, as well
as in comparison with the RADAU15, a traditional wide-spread FORTRAN implementation of the
Everhart integrator.

In addition to the $N$-body fitting, which requires a short-term $N$-body integration,
\PlP can perform the traditional long-term numeric integration. The integration scheme is
the same for both cases~--- it is the one based on $8$ Gauss--Legendre nodes. The
difference is in the step-size controlling: for short-term integrations we use a variable
step-size (aimed to achieve the maximum performance), while for long-term integrations we
use constant step (aimed to preserve the symplectic property).

\section{Statistical issues: analytical methods}
\label{sec_stat}
Statistics is an important component of \PlP. It includes some theoretical results
(classic and recent ones), as well as tools for numerical Monte Carlo simulations. The
newly-developed statistical theory implemented in \PlP mainly concerns the significance
levels of the periodograms. \PlP calculates the false alarm probability ($\FAP$) of
individual periodogram peaks using the method explained in
\citep{Baluev08a,Baluev08b,Baluev09a}, which is based on theory of extreme values of
random processes (the generalized Rice method). For a periodogram where the signal is
modelled by a trigonometric polynomial of degree $n$, the main $\FAP$ estimation formula
derived in the works by \citet{Baluev08a,Baluev09a} looks like
\begin{eqnarray}
\FAP(z) \lesssim M(z) \simeq W \alpha_n e^{-z} z^{n-1/2}, \nonumber\\
\alpha_n = \frac{2^n}{(2n-1)!!} \sum_{k=1}^n \frac{(-1)^{n-k} k^{2n+1}}{(n+k)!(n-k)!}, \nonumber\\
W = \Delta f T_\mathrm{eff}.
\label{prdg_fap}
\end{eqnarray}
where $z$ is the observed maximum peak on th periodogram, $\Delta f$ is the width of the
frequency band, and $T_\mathrm{eff}$ is the effective length of the time series. The
latter quantity is defined as $\sqrt{4\pi \disp t}$, where $\disp t$ is the weighted
variance of the times $t_{ji}$ (with the weights taken as $1/\sigma_{ji}^2$ at the
best-fit $\vec p$). This effective length is usually close to the plain time span of the
time series. The sign `$\lesssim$' in~(\ref{prdg_fap}) means that $M(z)$ represents an
upper bound for $\FAP(z)$ and simultaneously an asymptotic approximation for $\FAP(z)$
when $z\to\infty$. The approximation for the function $M(z)$ in~(\ref{prdg_fap}) was
obtained using the so-called ``assumption of the uniform phase coverage''. Regardless of
so apparently restrictive name, for the stated $\FAP$ evaluation task this assumption
works well in the majority of the practical cases, as we have shown, even for ultimately
strong spectral leakage (aliasing).

Strictly speaking, the formula~(\ref{prdg_fap}) was derived for the case when the RV
models are linear (except for the frequency parameter), and the noise uncertainties are
known a priori (no noise models involved). However, for more practical cases, including
weakly non-linear models and parametrized noise, the same formulae can be used in the
asymptotic sense for $N\to\infty$. See \citep{Baluev08b,Baluev13b} for the details.
Unfortunately, for the periodograms involving models with correlated noise of
Section~\ref{sec_rednoise}, we have not yet developed a reliable theory of the
significance levels. In this case \PlP will evaluate an approximation of the $\FAP$
according to some suggestive generalization of~(\ref{prdg_fap}) to the red-noise models,
but at present Monte Carlo simulations must be considered superior in this case.

\PlP is tuned to utilize the likelihood-ratio test for comparison of nested models. Given
two rival RV models: a base (more simple one) $\mu_\mathcal{H}$ and an alternative (more
complicated one) $\mu_\mathcal{K}$, we have the classical hypothesis testing task: is the
base hypothesis $\mathcal H$ consistent with the data, or it should be rejected in favour
of its alternative $\mathcal K$? This question can be answered after calculation of the
classic likelihood-ratio statistic
\begin{eqnarray}
Z = \ln \mathcal L_\mathcal{K}^* - \ln \mathcal L_\mathcal{H}^*, \nonumber\\
 \mathcal L_\mathcal{H}^* = \max_{\vec\xi_\mathcal{H}} \mathcal L_\mathcal{H}(\vec\xi_\mathcal{H}), \quad
 \mathcal L_\mathcal{K}^* = \max_{\vec\xi_\mathcal{K}} \mathcal L_\mathcal{K}(\vec\xi_\mathcal{K}).
\label{likrat}
\end{eqnarray}
The larger is $Z$, the greater is the observable advantage of $\mathcal K$ over $\mathcal
H$. \PlP, however, should honour the bias-reducing modification~(\ref{loglikmod}), which
leads to a modified likelihood-ratio by \citet{Baluev08b}, which is defined as
\begin{eqnarray}
\tilde Z = \frac{N_\mathcal{K}}{N}\left( \ln\tilde{\mathcal L}_\mathcal{K}^* - \ln\tilde{\mathcal L}_\mathcal{H}^* \right) +
 \frac{N_\mathcal{K}}{2} \ln \frac{N_\mathcal{H}}{N_\mathcal{K}}, \nonumber\\
 \tilde{\mathcal L}_\mathcal{H}^* = \max_{\vec\xi_\mathcal{H}} \tilde{\mathcal L}_\mathcal{H}(\vec\xi_\mathcal{H}), \quad
 \tilde{\mathcal L}_\mathcal{K}^* = \max_{\vec\xi_\mathcal{K}} \tilde{\mathcal L}_\mathcal{K}(\vec\xi_\mathcal{K}).
\label{likratmod}
\end{eqnarray}
where $N_{\mathcal H,\mathcal K} = N - d_{\mathcal H,\mathcal K}$ with $d_{\mathcal H,
\mathcal K}$ being the numbers of the degrees of freedom in the RV models to compare.

The quantity $\tilde Z$ represents a critical quantity for the decision: the larger is
$\tilde Z$, the less likely is $\mathcal H$ in comparison with $\mathcal K$. When the RV
models are linearisable, the asymptotic distribution of $2\tilde Z$ (for $N\to\infty$) is
the $\chi^2$-distribution with $d = d_\mathcal{K} - d_\mathcal{H}$ degrees of freedom
(under $\mathcal H$). This framework is used in \PlP to define the generalized
periodograms (Section~\ref{sec_prdg}) and the asymptotic confidence regions
(Section~\ref{sec_confreg}). In practice, at least for the confidence regions
determination task, the asymptotic $\chi^2$ distribution may work well, even when the RV
model is pretty complicated and non-linear \citep{Baluev13a}. For the periodograms we
however should use the formulae~(\ref{prdg_fap}) and the related statistical theory,
rather than the classical $\chi^2$ distribution. This is because the models involved in
the periodogram definition are not entirely linearisable \citep{Baluev13b}.

The definition~(\ref{likratmod}) differs from the classic one in~(\ref{likrat}) in the
normalization and offset which were introduced to compensate for the corrector $\gamma$
in~(\ref{loglikmod}). This $\gamma$ is different for the model $\mathcal H$ or $\mathcal
K$, so we needed to introduce the bias of $(N_\mathcal{K}/2) \ln
(N_\mathcal{H}/N_\mathcal{K}) \simeq d/2$ to make $\tilde Z$ asymptotically equivalent to
$Z$ (with a possible residual error of ${\sim} 1/N$). The normalizing factor
$N_\mathcal{K}/N$ does not alter the asymptotic properties of $\tilde Z$ and it has only
rather cosmetic purpose: it was chosen so that for the multiplicative noise model,
$\sigma_i^2 = \kappa/w_i$ with fixed weights $w_i$, the statistic $\tilde Z$ appears equal
to the statistic $z_3$ from \citep{Baluev08a}.

It is important that the model $\mathcal{K}$ includes $\mathcal{H}$ as a partial case or a
subset of lesser dimension, i.e. these models are nested. This implies, in particular,
that $d_\mathcal{H} < d_\mathcal{K}$ and the fit parameters of $\vec\xi_\mathcal{H}$
represent a subset of $\vec\xi_\mathcal{K}$.

Another small but useful statistical method, implemented in \PlP, is the Vuong test for
the comparison of non-nested rival models \citep{Baluev12}. It can be used to resolve the
period ambiguity due to the aliasing, or other types of ambiguity involving peer
(non-nested) models.

\section{Statistical issues: simulations}
\label{sec_simul}
For more intricate statistical tasks, \PlP allows to perform numerical Monte Carlo
simulations in a user-friendly manner. There are a few Monte Carlo algorithms that are
implemented in \PlP.

\subsection{Plain Monte Carlo assuming Gaussian noise and a single nominal model}
\label{subsec_MC}
This is a classical Monte Carlo scheme, which is used to model the distribution function
of the statistic $\tilde Z$ or the probability density of the best-fit estimations
$\vec\xi^*$. In this algorithm we assume that the true values of the parameters are more
or less known. The relevant simulated distribution functions, $P(\tilde Z |
\hat{\vec\xi})$ and $p(\vec\xi^* | \hat{\vec\xi})$ depend on the assumed nominal values
$\hat{\vec\xi}$, which a considered as true.
\begin{enumerate}
\item First of all, select some ``nominal'' (assumed ``true'') values $\hat{\vec\xi}$
somewhere in the region of interest. We may select e.g.\ the best-fitting model for this
goal, although such choice is not mandatory.

\item Given the chosen nominal vector $\hat{\vec\xi}$, evaluate the implied nominal RV
values and the RV errors variances $\sigma_i$ (or, for the red-noise framework, the full
noise covariance matrix).

\item Construct a simulated RV dataset by means of adding to the nominal RV curve the
simulated Gaussian errors, generated on the basis of previously evaluated uncertainties
and correlations.

\item Based on the simulated dataset, evaluate the value of the likelihood function at the
nominal parameter values from step~1, and the maximum value of $\tilde{\mathcal L}$ for
this trial. Based on these two values, evaluate the necessary modified likelihood ratio
statistic $\tilde Z$ for this trial.

\item Save this value of $\tilde Z$, as well as the set of the simulated best fitting
parameters (when necessary), and return to step~3, if the desired number of trials has not
been accumulated yet.
\end{enumerate}
Therefore, this classical method is not self-closed: we should feed the simulation with
the nominal vector $\hat{\vec\xi}$. Due to this weakness, the results of the simulation
are usable when the functions $P(\tilde Z | \hat{\vec\xi})$ and $p(\vec\xi^* |
\hat{\vec\xi})$ do not demonstrate large dependence on $\hat{\vec\xi}$ (at least for
the expected realistic values of $\hat{\vec\xi}$, e.g.\ the ones covering the uncertainty
region).

We actually recommend this simulation in practice only to detect a significant
non-linearity of the task specified, or, vice versa, to show that a particular task can be
dealt with by means of the linear asymptotic methods discussed in previous sections. This
is achieved by means of the comparison of the asymptotic confidence regions or of the
asymptotic $\chi^2$ likelihood-ratio distribution with the results of simulations. We must
note, that e.g.\ non-elliptic shape of the parametric confidence regions (asymptotic or
Monte Carlo ones) does not yet imply any genuine non-linearity at all. Although such a
deviation from ellipticity is usually deemed as an indicator of the non-linearily, it may
be often caused by other reasons like, e.g.\ an uncareful choice of the parametrization
\citep{Baluev13a}. To check that the non-linearity is indeed genuine (``endogeneous'') and
that it really requests the use of complicated non-asymptotic treatment, it is necessary
to verify the agreement between the asymptotic results and the results of the classic
Monte Carlo simulation.

When the model is proven to be significantly non-linear, the classic Monte Carlo scheme
does not offer more realistic confidence regions or confidence probabilities, so it should
not be used for this goal either. The naive interpretation of the classic Monte Carlo
results leads to some caveats. For example, it may double the statistical bias of the
maximum-likelihood estimations, rather than to compensate it: the Monte Carlo trials will
generate ``mock'' best-fit parametric solutions that are biased relatively to the actual
best-fit one in the same manner as this actual best-fit configuration is biased relatively
to the truth.

\subsection{Bootstrap simulation}
\label{subsec_bstrp}
The bootstrap is used when there is a danger that the RV errors are not really Gaussian,
although we must note that the actual practical profit from the bootstrap in the exoplanet
searches still remains without detailed investigation.
\begin{enumerate}
\item Evaluate the best fitting model and the resulting RV residuals.
\item Apply random shuffling procedure of the residuals (we do this separately to each
sub-dataset of our combined time series).
\item Evaluate the statistic $\tilde Z$ and best fitting parameters in the same manner as
in the plain Monte Carlo simulation.
\item Save the resulting value of $\tilde Z$ and parameters and return to step~2.
\end{enumerate}
The bootstrap shares all weaknesses of the classical Monte Carlo scheme, except for the
assumption of the noise Gaussianity. In addition, it possesses extra disadvantages, for
example it is meaningful only with a white-noise RV model, because random shuffling of the
residuals basically destroys any correlational structure of the RV noise, which a
red-noise model tries to deal with.

Another weakness of the bootstrap simulation is that it does not work well for the noise
parameters \citep{Baluev13a}. Their bootstrap-simulated values are concentrated in an
unexpectedly small region, much smaller than the real uncertainty domain. Consequently,
the result of such a simulation looks roughly as if these noise parameters were held fixed
during the simulation. This also results in a different and rather unpredictable behaviour
of the statistic $\tilde Z$ obtained using such a simulation.

\subsection{Genuinely frequentist Monte Carlo simulation}
\label{subsec_FMC}
The existing criticism of the above-described Monte Carlo algorithms is due to their
sensitivity to some assumed constant ``true'' or ``nominal'' vector of the fit parameters.
This is one of the main arguments of many statisticians, which they often use to highlight
the advantages of the Bayesian approach. The Bayesian methods do not rely on a single
nominal vector: instead, they deal with a scattered prior distributions covering a large
parametric domain.

However, it appears that this led us to an unjustified opposing the Bayesian methods with
the frequentist ones. Although the above simulation schemes do suffer from the issue of
constant nominal values, this issue can be eliminated in the genuinely frequentist
framework. Leaving aside the philosophy, the main \emph{techincal} difference between the
Bayesian and frequentist methods is in how they treat the uncertainty of the nominal
values of the fit parameter. While the Bayesian methods use weighted averaging with some
pre-set prior probability density, the genuine frequentist approach is based on the
worst-case principle, and uses the maximization or minimization in place of the averaging.
If in the simplified frequentist approach we dealt with the distribution function
$P(\tilde Z | \hat{\vec\xi})$, in the genuine frequentist treatment we should replace it
with
\begin{equation}
P_\mathrm{worst}(\tilde Z) = \min_{\vec\xi \in \Xi} P(\tilde Z | \vec\xi),
\label{Pw}
\end{equation}
which means the worst-case confidence probability. Standing on the Bayesian positions with
probabilistic $\vec\xi$, the distribution of $\tilde Z$ would be expressed as
\begin{equation}
P_\mathrm{mean}(\tilde Z) = \int\limits_{\Xi} P(\tilde Z | \vec\xi) p(\vec\xi) d\vec\xi
\label{Pm}
\end{equation}
with $p(\vec\xi)$ being the prior distribution of the parameters. Obviously, the
difference between~(\ref{Pw}) and~(\ref{Pm}) is crucial, but we cannot say that one of
them is generally better, or on contrary deprecated. Each approach has its own advantages
and disadvantages in concrete special circumctances; some of them are briefly discussed in
\citep{Baluev13a}. Obviously, in the frequentist approach we need only to outline a
parametric domain $\Xi$, and any prior density inside this domain does not play any role
when we find the minimum.

The entire algorithm of the genuine frequentist simulation would look as the following:
\begin{enumerate}
\item Select an $i$th trial point $\hat{\vec\xi}_i$ (possibly, residing inside a given
parametric domain $\Xi$).
\item Run the plain Monte Carlo algorithm of Sect.~\ref{subsec_MC} assuming that the true
parameters correspond to the selected point.
\item Save the simulated distribution $P(\tilde Z | \hat{\vec\xi}_i)$ of the test
statistic of interest ($\tilde Z$ in our case) and return to step~1.
\item When a sufficiently dense coverage of the mentioned in step~1 parametric domain is
reached, evaluate the function $P_\mathrm{worst}(\tilde Z) = \min_i P(\tilde Z |
\hat{\vec\xi}_i)$.
\end{enumerate}
After that, the rigorous frequentist false alarm probability associated with an
\emph{observed} value $\tilde Z_*$ (which was obtained using exactly the same models that
were used during the simulation) can be calculated as $1-P_\mathrm{worst}(\tilde Z_*)$.

At present, \PlP does not incorporate Bayesian tools, but the genuine frequentist
simulation can be organized by means of calling it subsequently from an external shell
script. To do this we should first generate some set of $\hat{\vec\xi}_i$, saving it in a
file. This can be done with \PlP by means of the plain Monte Carlo algorithm, or using
another preferred external procedure. After that, \PlP can be subsequently executed for
each saved $\hat{\vec\xi}_i$ to perform the simulation of step~2, saving the relevant
distribution $P(\tilde Z | \hat{\vec\xi}_i)$. Then these distributions should be processed
externally to generate $P_\mathrm{worst}(\tilde Z)$. This is exactly how we estimated the
significance of the planet GJ~581~\emph{e} in \citep{Baluev13a}.

\section{Conclusions}
\label{sec_conc}
We have a hope that \PlP functionality will grow further in future, not limited by the
things that we have described in the paper. In particular, it would be tempting to add
some algorithms of Bayesian simulations, and to have some capabilities of dealing with
astrometric data, because of the forthcoming domination of GAIA astrometry. Among more
technical things, we would like to make \PlP able to work in a multi-threaded mode,
profiting from the full capabilities of modern multi-core CPUs or even from GPU computing
(at present, \PlP is single-threaded).

\PlP is a free and open-source software. We do not set any limitation on the use of itself
or of its source code (except for providing a proper reference to the present paper).
Anyone who is interested is allowed to freely modify its code to improve it or to adapt it
to their specific needs, although it would be of course preferrable to incorporate a
significant and worthy improvement in \PlP itself rather than to make an independent fork.

\section*{Acknowledgements}
This work was supported by Russian Foundation for Basic Research (project 12-02-31119
mol\_a) and by the programme of the Presidium of Russian Academy of Sciences
``Non-stationary phenomena in the objects of the Universe''. I would like to express the
gratitude to my collegue, Dr. I.I.~Nikiforov, as well as to the reviewers, for providing
useful comments during preparation of this manuscript. Also, we acknowledge that a few of
linear algebra algorithms used in \PlP represent re-worked versions of the relevant GNU
Scientific library subroutines.

\bibliographystyle{model2-names}
\bibliography{PlanetPack}

%% not want to use model2-names.bst.



%% \bibitem must have one of the following forms:
%%   \bibitem[Jones et al.(1990)]{key}...
%%   \bibitem[Jones et al.(1990)Jones, Baker, and Williams]{key}...
%%   \bibitem[Jones et al., 1990]{key}...
%%   \bibitem[\protect\citeauthoryear{Jones, Baker, and Williams}{Jones
%%       et al.}{1990}]{key}...
%%   \bibitem[\protect\citeauthoryear{Jones et al.}{1990}]{key}...
%%   \bibitem[\protect\astroncite{Jones et al.}{1990}]{key}...
%%   \bibitem[\protect\citename{Jones et al., }1990]{key}...
%%   \harvarditem[Jones et al.]{Jones, Baker, and Williams}{1990}{key}...
%%


% \end{thebibliography}

\end{document}